\documentclass{jetpl}
\twocolumn

\usepackage{slashed}
\usepackage[dvips]{color}
\usepackage[dvips]{epsfig}
\usepackage{latexsym}
\usepackage{bm}
\usepackage{upgreek}
\usepackage{mathrsfs}
\usepackage{times}
\usepackage{amsthm}
\usepackage{amssymb}
\usepackage{epsfig}
\usepackage{graphicx}
\usepackage{amsmath}

\usepackage{color}

\definecolor{purple}{rgb}{0.8,0,0.6}

\title{Emergent Weyl fermions and the origin of $i=\sqrt{-1}$ in quantum mechanics
}

\rtitle{Emergent Weyl fermions and the origin of $i=\sqrt{-1}$ in quantum mechanics
}

\sodtitle{Emergent Weyl fermions and the origin of $i=\sqrt{-1}$ in quantum mechanics
}

\author{G.E. Volovik $^{*+}$
\/\thanks{e-mail: volovik@boojum.hut.fi}
and M.A. Zubkov $^{**++}$\/\thanks{e-mail: zubkov@itep.ru}
}

\rauthor{G.E.Volovik, M.A.Zubkov}

\sodauthor{Volovik, Zubkov}

\address{
$^{*}$ O.V. Lounasmaa Laboratory, Aalto University, School of Science and
Technology, P.O. Box 15100, FI-00076 AALTO, Finland
\\
$^+$ Landau Institute for Theoretical Physics RAS, Kosygina 2,
119334 Moscow, Russia
\\
$^{**}$ ITEP, B.Cheremushkinskaya 25, Moscow, 117259, Russia
\\
$^{++}$ The University of Western Ontario, Department of Applied
Mathematics, 1151 Richmond St. N., London (ON), Canada N6A 5B7
  }

\dates{\today}{*}


\abstract{
Conventional quantum mechanics is described in terms of complex numbers. However, all physical quantities are real. This indicates, that the appearance of complex numbers in quantum mechanics may be the emergent phenomenon, i.e. complex numbers appear in the low energy description of the underlined high energy theory. We suggest a possible explanation of how this may occur. Namely, we consider the system of multi - component Majorana fermions. There is a natural description of this system in terms of real numbers only. In the vicinity of the topologically protected Fermi point this system is described by the effective low energy theory with Weyl fermions. These Weyl fermions interact with the emergent gauge field and the emergent gravitational field.
}

\begin{document}

\maketitle


Properties of topological media (topological insulators, topological superconductors, Weyl semi-metals, etc. \cite{HasanKane2010,QiZhang2010})   are generic. They are protected by  topology and thus are robust to deformation of the system and not sensitive to the detailed microscopic (atomic) structure of a given topological material.  These systems experience relativistic phenomena at low energy, which earlier have been the prerogative of particle physics. The emergence of Lorentz invariance and other physical laws and their topological stability  suggest  that Standard Model of particle physics and  Einstein theory of gravitational field may have the status of effective theories. The chiral elementary particles (quarks and leptons), gauge and Higgs bosons, and the dynamical tetrad field representing gravity may naturally emerge in the low-energy corner of the quantum vacuum, provided the vacuum has  Weyl points (topologically protected point nodes in the energy spectrum
 \cite{Froggatt1991,Volovik2003,Horava2005}).

Particle physics and physics of many-body systems have many common properties.
Last decades it became clear that the close connection between the two areas
arises from the common topology. In both systems the main role is played by fermionic particles:
quasiparticles in condensed matter and quarks and leptons in the Standard Model.
At low temperature the gapped (massive) fermions are frozen out and only gapless (massless) quasiparticles may survive. The gaplessness is the fragile property, since it can be violated by interaction between fermions. Nevertheless, there exist fermionic systems, in which the  gaplessness (masslessness) is robust to interaction.
These are topological materials, where stability of nodes in the energy spectrum is protected by the conservation of topological invariants of different types \cite{Horava2005}.
The most familiar example of topologically protected zeroes in fermionic spectrum is the
Fermi surface in metals: it survives even if the interaction becomes so strong that quasiparticles are not any more well defined  \cite{Volovik2003}.

The most close connection to particle physics comes from the condensed matter systems with point nodes.
The spectrum of fermions in the vicinity of point nodes typically acquires the relativistic form. Moreover, for these low-energy fermions the analog of Lorentz symmetry emerges together
with effective gauge and gravity fields; the quasiparticles live in effective curved space - time, whose geometry is formed by certain collective excitations of the microscopic system
\cite{Froggatt1991,Volovik2003,Volovik1986A,Volovik2011}.
The topologically protected point  nodes exist in 3+1 superfluid $^3$He-A \cite{Volovik2003}.
Near these points the quasiparticles are similar to the right-handed and the left-handed electrons of the Standard Model obeying Weyl equation. The Weyl points are suggested to exist in 3+1 Weyl semimetals
\cite{Abrikosov1971,Abrikosov1998,Burkov2011,XiangangWan2011,Weylsemimetal}.  Dirac point nodes
are well known in 2+1 graphene
\cite{Semenoff:1984dq}. The point nodes emerge also in the spectrum of
fermionic modes on the surface and interfaces of the fully gapped topological insulators
\cite{VolkovPankratov1985,HasanKane2010,Xiao-LiangQi2011} and topological superfuids
\cite{SalomaaVolovik1988,Volovik2009}.
The Weyl and Dirac points represent  the exceptional (conical, diabolic) points of level crossing, which avoid the level repulsion  \cite{NeumannWigner}. Topological invariants for such points, at which the branches of spectrum merge, were introduced by Novikov \cite{Novikov1981}.
In our case the crossing points occur in momentum space \cite{Avron1983,Volovik1987}.

\begin{figure}
\includegraphics[width=7cm]{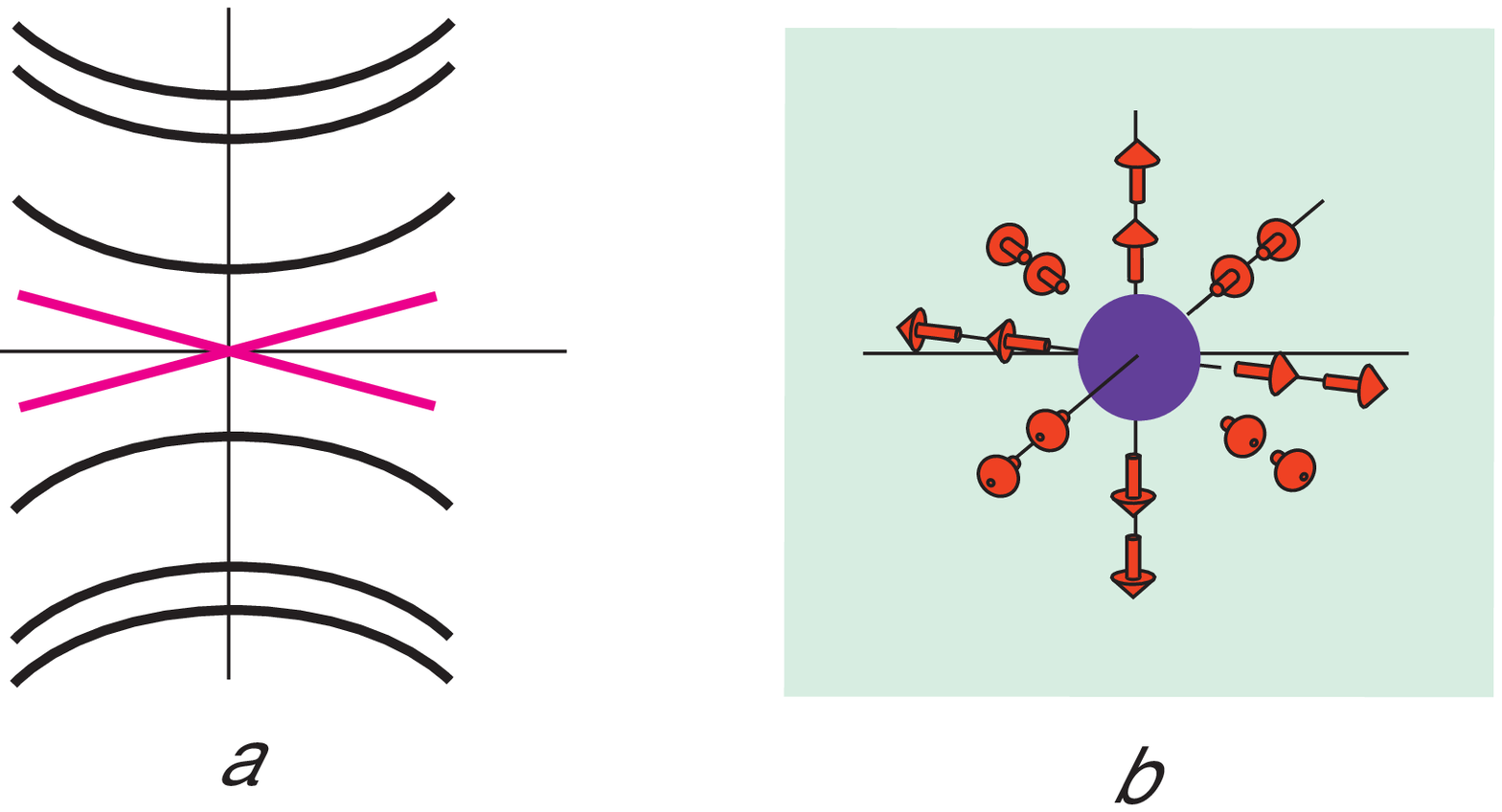}
\caption{(a) Various branches $E_k$ of the spectrum for the microscopic hamiltonian are represented schematically. Two branches $E_1, E_2$ cross each other at the position of the Fermi point. The other spectrum branches correspond to massive excitations, that decouple at low energies.
(b) In this figure the pattern of $m^L({\cal P})$ around the position of the branches crossing point ${\cal P}^{(0)}$ is represented. This pattern corresponds to the value of the topological invariant (\ref{A(K)-expansion00}) equal to $N = +1$. This topological invariant protects the branches crossing from the disappearance.
}
\label{WeylFermions}
\end{figure}

In the vicinity of the Weyl point the system is effectively
described by the two - component fermion field $\Psi$ obeying Weyl equation
\cite{Froggatt1991,Horava2005,Volovik1986A}
\begin{eqnarray}
{\cal D}  \Psi &=& 0 \label{D} \\ {\cal D} & = & \frac{1}{2}\Bigl(e^{\mu}_a \,
\sigma^a \left(i \partial_\mu -p^{(0)}_\mu\right) +
\sigma^a \left(i \partial_\mu -p^{(0)}_\mu\right)\, e^{\mu}_a \Bigr) \,\nonumber
\end{eqnarray}
Here  $p^{(0)}_\mu$ is the position of the Fermi point, whose space - time variation
 gives rise to the effective dynamical electromagnetic field $F_{\mu\nu}=\partial_\mu p^{(0)}_\nu
-  \partial_\nu p^{(0)}_\mu$;
and $e^{\mu}_a$ is an emergent vierbein, which describes the effective space-time with metric $g^{\mu\nu}=\eta^{ab}e^{\mu}_a e^{\nu}_b$  (we denote metric of Minkowsky space - time by $\eta^{ab}$). In Eq. (\ref{D}) the vierbein does not commute with derivative. The latter does not contain spin connection. This means that we deal with the effective gravity with torsion in the limit of vanishing $SO(3,1)$ curvature. This is the so - called teleparallel gravity \cite{teleparallel}. The corresponding geometry differs both from Riemann - Cartan and Riemannian geometry. In superfluid $^3$He-A  equation  (\ref{D})  has been explicitly obtained by expansion
of the Bogoliubov-de Gennes Hamiltonian near the Weyl point
\cite{Volovik1986A};  for the expansion near Dirac point in $2+1$ D graphene see Ref. \cite{VK2010,Manes2013,ZV2013}.
In both cases the complicated atomic structure of liquid and solid is reduced to
the description in terms of the effective two-component spinors.

The emergence of Weyl spinor has important consequences both in the
condensed matter physics and in the high energy physics. This is because
 Weyl fermions represent the building blocks of the Standard Model of particle physics (SM).
 Emergence of Weyl fermions in condensed matter together with Lorentz invariance, effective gravity and gauge fields and the topological stability of emergent phenomena suggest  that  SM  and  Einstein theory of gravitational field (GR) may have the status of effective theories. The chiral elementary particles (quarks and leptons), gauge and Higgs bosons, and the dynamical vierbein field may naturally emerge in the low-energy corner of the quantum vacuum, provided the vacuum has topologically protected Weyl points.

In this approach it is assumed from the very beginning, that the underlying microscopic
 Hamiltonian is the complex-valued
$N\times N$ matrix, acting on the complex valued $N$-component wave function.
At the first glance this is not surprizing, since quantum mechanics (or, in general, the quantum field theory), is described by the language of complex numbers. All the equations of quantum mechanics  --  the Weyl, Dirac and Schr\"odinger equations -- are described in terms of complex numbers.
On the other hand, the imaginary unit $i=\sqrt{-1}$ is the product of human mind,
which is mathematically convenient, while all the measured physical
quantities are real. This may imply that the imaginary unit $i=\sqrt{-1}$
should not enter any physical equation.  It is known that Schr\"odinger strongly resisted to introduce
$i$ into his wave equations (see Yang \cite{Yang}).  The  Schr\"odinger  equation is obtained
from Dirac equation in the non-relativistic limit, when the characterisic non-relativistic energy $E\ll mc^2$,
where $m$ is the mass of Dirac particle. In turn, the Dirac equation is obtained from Weyl equation, when the
two-component left-handed and the two-component right-handed Weyl fermions are mixed to form the 4-component Dirac particle.

The complex numbers may be represented by $2\times 2$ real - valued matrices
\begin{equation}
q=a+ib\rightarrow Q=a \times \hat{1} +\hat{i}_{\rm eff}\times b =\left(\begin{array}{cc}a&  -b\\
b&a\end{array}\right)~,
\label{-1element}
\end{equation}
where matrices of real and imaginary units are $\hat{1}$ and $\hat{i}_{\rm eff}$
\begin{equation}
\hat{1}=\left(\begin{array}{cc}1& 0\\
0&1\end{array}\right),~~\hat{i}_{\rm eff}=\left(\begin{array}{cc}0& -1\\
1&0\end{array}\right),~~\hat{i}_{\rm eff}^2=-1
\label{Units}
\end{equation}
The complex conjugation is defined by matrix $\hat{C}$, which can be chosen as:
\begin{eqnarray}
\hat{C}&=&\left(\begin{array}{cc}1&  0\\ 0&-1\end{array}\right)~~,~~ Q^*\equiv C Q C =a\times \hat{1}-b \times \hat{i}_{\rm eff}\nonumber\\ && q^*=a-ib
\label{ComplexConjugation}
\end{eqnarray}

It is necessary to find out the reason, why such structure entered
the wave function in quantum mechanics, i.e. what is the geometric, symmetry
or topological origin of the algebra of complex numbers in the formalism of
quantum mechanics.
One scheme  suggested by Adler \cite{Adler2004}
is the so-called trace dynamics, where  $g=i_{\rm eff}$  appears as an anti-self-adjoint operator:
$i_{\rm eff}=-i_{\rm eff}^\dagger$, $i_{\rm eff}^2=-1$.

The general scheme that describes how Weyl spinors appear in the low energy approximation to the system of multi - component fermions was given by Ho\v{r}ava  \cite{Horava2005} and developed in \cite{VZ2014_NPB}. Here we concentrate on  the  underlying microscopic theory, which is described solely in terms of the real numbers, i.e. in terms of Majorana fermions. It is shown, that there exists the special choice of coordinates, in which the low energy description is again given in terms of Weyl spinors. This scheme demonstrates that the effective imaginary unit $i_{\rm eff}$ may be also
the emergent phenomenon in quantum mechanics, and this $i_{\rm eff}$ emerges together with Weyl fermions, gauge fields,  gravity, and observed coordinate space itself.

Majorana fermions are described by the $N$-component real - valued spinor wave function $\chi({\bf Q})$. We start from the functions defined in the coordinates ${\bf Q}$. These coordinates not necessarily coincide with the coordinates of observed space, in which the low energy physics is described in terms of Weyl fermions. The evolution in time of the wave function is given by the equation
$
\partial_t \chi = \hat{A}\chi,
$
where $A$ is the operator. This operator may be non - local in the original coordinates $\bf Q$. In the language of quantum field theory the dynamics of Majorana fermions is governed by the functional integral over fictitious $N$ - component anti - commuting variables $\psi$ (Grassmann variables). The partition function of the theory has the form:
\begin{equation}
Z = \int D \psi {\rm exp}\Bigl( -\int d t \sum_{{\bf Q}} \psi_{\bf Q}^T(t)
(\partial_t + \hat{A}) \psi_{\bf Q}(t) \Bigr)\label{FI}
\end{equation}
The conventional form of the functional integral for Majorana fermions is recovered, when operator $\hat{A}$ is identified with  the Hamiltonian multiplied by $i$. However, written in the form of Eq. (\ref{FI}) the functional integral does not contain imaginary unity. Besides, it contains only one $N$ - component Grassmann spinor $\psi$. (The functional integral for usual fermions would contain the second independent integration variable $\bar{\psi}$ that corresponds to the hermitian conjugated field.) In Eq. (\ref{FI}) there is the sum over the initial coordinates $\bf Q$. Later we shall imply that these coordinates belong to ${\bf R}^3$, so that the sum over these coordinates should be understood as an integral over $d^3 {\bf Q}$. However, in principle, the theory may be developed in such a way, that these coordinates are discrete.

In the vicinity of the Fermi - point the system of multi - component Majorana fermions is described effectively by the model with the reduced number of fermion components $N_{\rm reduced}$. $N_{\rm reduced}$ is the minimal number allowed by momentum space topology. The proof \cite{VZ2014_NPB} is based on the general property of the Hermitian (and anti - Hermitian) operators: several branches of spectrum repel each other. This means that if the two branches cross each other at a point, any small perturbation pushes them apart from each other. That's why the crossing point survives only if it is protected by something that does not allow the spectrum branches to be repelled. Fermi point appears as the position of the branches crossing point for operator  $\hat A$.

Minimal number of the branches that cross each other allowed by nontrivial momentum space topology is two. This corresponds to the number of the reduced spinor components equal to four. More explicitly, there exists the orthogonal operator $\hat{\Omega}$ that brings operator $\hat A$ to the form of the block - diagonal $N\times N$ matrix  ${A}^{{\rm block}\,{\rm diagonal}} = {\rm diag}\Bigl( E_1({\cal P})\hat{i}_{\rm eff}, E_2({\cal P})\hat{i}_{\rm eff}, E_3({\cal P})\hat{i}_{\rm eff}, ... \Bigr)$. This matrix depends on the new coordinates $\cal P$. These coordinates parametrize the spectrum branches of operator $\hat A$. We refer to them as to the coordinates in generalized momentum space. The first $4\times 4$ block of this matrix has the form ${A}^{{\rm block}\,{\rm diagonal}}_{\rm reduced} = {\rm diag}\Bigl( E_1({\cal P})\hat{i}_{\rm eff} , E_2({\cal P})\hat{i}_{\rm eff}\Bigr)$ with some functions $E_1({\cal P})$ and $E_2({\cal P})$ that coincide at  ${\cal P} = {\cal P}^{(0)}$.
The remaining block of matrix ${A}^{{\rm block}\,{\rm diagonal}}_{\rm massive}$ corresponds to the
"massive" branches. The functional integral can be represented as the
product of the functional integral over "massive" modes and the integral
over $4$ reduced fermion components. The Fermi point appears at ${\cal P}^{(0)}$ if chemical potential is equal
to the value $E_{1,2}({\cal P}^{(0)})$. Then the four reduced components dominate the functional integral while the remaining "massive" components decouple and do not influence the dynamics. This pattern in illustrated by FIG. 1 (a).

The possible explanation why the value of chemical potential is equal to the value $E_{1,2}({\cal P}^{(0)})$ is that during inflation the chemical potential is self - tuned in such a way, that it becomes equal to the energy level crossing point (see,  for example, the arguments in favor of this supposition in \cite{Froggatt1991}). However, this question remains out of the scope of the present paper, and we simply postulate, that the chemical potential is equal to $E_{1,2}({\cal P}^{(0)})$.

Functions $E_{1,2}({\cal P})$ may be non - analytical at ${\cal P}^{(0)}$. However, the block - diagonal form ${\rm diag}\Bigl( E_1({\cal P})\hat{i}_{\rm eff} , E_2({\cal P})\hat{i}_{\rm eff}\Bigr)$ of the reduced matrix is related by the $4\times 4$ orthogonal transformation  (that commutes with $\hat{i}_{\rm eff}$) to the $4\times 4$ matrix $A_{\rm reduced}({\cal P})$ of a more general form.  The dependence of $A_{\rm reduced}({\cal P})$ on $\cal P$ is analytical (except for the marginal cases). This is typical for the functions that are encountered in physics. The non - analytical functions represent the set of vanishing measure in space of functions.

Let us denote the components of the reduced $4$ - component spinors by $
\psi = \Bigl(\psi^1,  \psi^2,  \psi^3,  \psi^4 \Bigr)^T$.
$A_{\rm reduced}({\cal P})$ commutes with $\hat{i}_{\rm eff}$. This allows to introduce the new two - component spinors
\begin{equation}
\Psi_{\cal P} = \left(\begin{array}{c}\psi_{\cal P}^1 + i \psi_{\cal P}^2\\
\psi_{\cal P}^3 + i \psi_{\cal P}^4 \end{array} \right), \quad
\bar{\Psi}_{\cal P} = \left(\begin{array}{c}\psi_{\cal P}^1 - i \psi_{\cal P}^2 \\ \psi_{\cal P}^3 - i \psi_{\cal P}^4 \end{array}
\right)^T\label{barPsi0}
\end{equation}
and express the partition function in terms of the two spinors $\Psi$ and $\bar{\Psi}$:
\begin{eqnarray}
Z & = & \int D \Psi D\bar{\Psi} {\rm exp}\Bigl(\frac{i}{2} \int d t \sum_{{\cal P}} \Bigl[\bar{\Psi}_{{\cal
P}}(t) (i\partial_t - \hat H )\Psi_{\cal P}(t)\nonumber\\&&+(h.c.)\Bigr]  \Bigr),\quad \hat H  =   m^L_k( {{\cal P}}) \hat \sigma^k
+  m({{\cal P}})\label{Z__00}
\end{eqnarray}
We used that $\int dt (\psi_{\cal P}^1 - i\psi_{\cal P}^2)  \partial_t (\psi_{\cal P}^1 + i \psi_{\cal P}^2) = \int dt(\psi_{\cal P}^1,\psi_{\cal P}^2) \partial_t (\psi_{\cal P}^1,\psi_{\cal P}^2)^T$ (notice, that $\int dt \psi_{\cal P}^1  \partial_t  \psi_{\cal P}^2 = \int dt \psi_{\cal P}^2 \partial_t \psi_{\cal P}^1$). Besides, we substitute      $i \bar{\Psi}_{\cal P}  \hat H \Psi_{\cal P} = {\psi}_{\cal P}^T \hat A_{\rm reduced} \psi_{\cal P}$. This equality becomes obvious, when matrix $A_{\rm reduced}$ is block - diagonal, then we must check only, that
$i (\psi_{\cal P}^1 - i\psi_{\cal P}^2)  E_1({\cal P}) (\psi_{\cal P}^1 + i \psi_{\cal P}^2) = (\psi_{\cal P}^1,\psi_{\cal P}^2)i_{\rm eff}E_1({\cal P})(\psi_{\cal P}^1,\psi_{\cal P}^2)^T$ (notice that $\psi^1 \psi^1 = \psi^2 \psi^2 = 0$), and the same for the components $\psi^3,\psi^4$. Hamiltonian $H$ that is related in this way to $A_{\rm reduced}^{{\rm block}\,{\rm diagonal}}$ has the form $H = \frac{E_1+E_2}{2} + \frac{E_1 - E_2}{2}\sigma^3$. The $4\times 4$ orthogonal transformation that relates $A_{\rm reduced}^{{\rm block}\,{\rm diagonal}}$ with $A_{\rm reduced}$ and commutes with $i_{\rm eff}$ can be written as $\Omega = {\rm exp}\Bigl(1\otimes \omega_A + i_{\rm eff} \otimes \omega_S\Bigr)$, where $\omega_A$ is skew - symmetric while $\omega_S$ is symmetric. Being applied to $\Psi$ this transformation becomes $\tilde{\Omega} = {\rm exp}\Bigl(\omega_A + i \omega_S\Bigr)\in U(2)$ and results in $\Psi \rightarrow \tilde{\Omega}\Psi, \bar{\Psi}\rightarrow \bar{\Psi} \tilde{\Omega}^+$. That's why transformation $\hat A \rightarrow \Omega \hat A \Omega^T$ results in unitary transformation of Hamiltonian $\hat H \rightarrow  \tilde{\Omega} \hat H \tilde{\Omega}^+$. This proves, that for any $A_{\rm reduced}$ there exists the Hamiltonian $H$ of the form of Eq. (\ref{Z__00}) such that $i \bar{\Psi}_{\cal P}  \hat H \Psi_{\cal P} = {\psi}_{\cal P}^T \hat A_{\rm reduced} \psi_{\cal P}$.

Analytical functions $m^L_k, m$ vanish at the Fermi point ${\cal P}^{(0)}$. The position of the branches crossing is stable if it is protected by the nonzero value of the invariant:
\begin{equation}
N= \frac{\epsilon_{ijk}}{8\pi} ~
   \int_{\sigma}    dS^i
~\hat{ m}^L\cdot \left(\frac{\partial \hat{ m}^L}{\partial {\cal P}_j}
\times \frac{\partial \hat{ m}^L}{\partial {\cal P}_k} \right),
\label{N0}
\end{equation}
where $\hat{ m}^L = \frac{{ m}^L}{|{ m}^L|}$, while $\sigma$  is the $S^2$ surface around the point.
For $N=\pm 1$ in Eq.(\ref{N0}) the expansion near the
hedgehog point at ${\cal P}^{(0)}_j$ in $3D$ ${\cal P}$-space gives
\begin{equation}
m^L_i({\cal P})\approx f_i^j({\cal P}_j-{\cal P}^{(0)}_j),\,  m({\cal P})\approx f_0^j({\cal P}_j-{\cal P}^{(0)}_j)
\label{A(K)-expansion00}
\end{equation}
(We also expand in powers of ${\cal P}_j-{\cal P}^{(0)}_j$ the function $m({\cal P})$ that does not enter Eq. (\ref{N0}).)
This hedgehog point is illustrated by FIG.1 (b).

 Let us identify quantities $\cal P$ with the eigenvalues of operator $\hat {\cal P} = - {i}  \, \frac{\partial}{\partial {\bf X}}$, where by $\bf X$ we denote the new coordinates. We denote $\Psi_{\bf X} = \frac{1}{\sqrt{\cal V}}\sum_{\cal P} e^{i{\cal P}{\bf X}} \Psi_{\cal P}$ and $\bar{\Psi}_{\bf X} = \frac{1}{\sqrt{\cal V}}\sum_{\cal P} e^{-i{\cal P}{\bf X}}\bar{\Psi}_{\cal P}$, where $\cal V$ is the volume of $3$D space. $\Psi_{\bf X}$ and $\bar{\Psi}_{\bf X}$ are the new independent Grassmann variables. In general case coordinates $\bf X$  do not coincide with the original coordinates $\bf Q$. This means, that the fields local in $\bf Q$ are not local in coordinates $\bf X$ and vice versa.
 As a result the
partition function of the theory receives the form
$
Z = \int D \Psi D\bar{\Psi}  e^{i S[e^j_a, B_j,
\bar{\Psi},\Psi]}
$
with
\begin{eqnarray}
S =   \frac{1}{2} \Bigl(\int d t \,e\,  \sum_{{\bf X}}
\bar{\Psi}_{{\bf X}}(t)  e_a^j  \hat \sigma^a i \hat D_j  \Psi_{\bf X}(t) +
(h.c.)\Bigr)\label{Se0}
\end{eqnarray}
The sum is over $a,j = 0,1,2,3$ while $\sigma^0 \equiv 1$. We represented here $f_i^j = e \, e_i^j$. In  Eq. (\ref{Se0})  $\hat
D$ is the covariant derivative that includes the $U(1)$ gauge field $B = {\cal P}^{(0)}$. The given representation for $f_i^j$ is chosen in this way in
order to interpret the field $e_i^j$ as the vierbein. This means, that we
require
\begin{eqnarray}
e^0_a &=& 0, \, {\rm  for} \, a=1,2,3; \quad e \times e_0^0=1;\nonumber\\ e^{-1} &=&
e_0^0 \times {\rm det}_{3\times 3}\, e^i_a = e_0^0 =  {\rm det}_{4\times 4}
e^i_a
\end{eqnarray}

Now let us take into account the interactions between the
original Majorana fermions. They result in  the fluctuations of ${\cal P}^{(0)}_k$ and $e^j_k$. Besides, there may appear the emergent fields that interact with the reduced  two - component spinors $\Psi,\bar{\Psi}$ via the terms that break the fermion number conservation (those terms contain bilinear combinations $\Psi_A\Psi_B$ and $\bar{\Psi}^A\bar{\Psi}^B$). Here we do not consider such fermion number breaking interactions.  We assume, that there exists the mechanism, that suppresses them. In addition, we imply, that the fluctuations of ${\cal P}^{(0)}_k$ and $e^j_k$ are long - wave, so that
the corresponding variables should be considered as if they would not depend
on coordinates. The
partition function of the theory receives the form
$
Z = \int D \Psi D\bar{\Psi} D e^i_k D B_k e^{i S_0[e, B] + i S[e^j_a, B_j,
\bar{\Psi},\Psi]}
$
while in  Eq. (\ref{Se0})  $\hat
D$ is the covariant derivative that includes the $U(1)$ gauge field $B$ fluctuating around the position of the Fermi point.
$S_0[e, B]$ is the part of the effective action that depends on $e$ and $B$
only.
Both these fields represent certain collective excitations of the
microscopic theory. (It is assumed, that the value of the emergent
electromagnetic field is much larger than the order of magnitude of quantity
$|\nabla e^k_a|$.) We impose the antiperiodic boundary conditions in time on the spinor fields in the synchronous reference frame, where $\langle e^j_0\rangle = 0$ for $j=1,2,3$.

Eq. (\ref{Se0}) is the action for the two - component Weyl spinor in the presence of the gravitational field given by the vierbein $e^j_k$ and the Electromagnetic field $B_k$. Weyl equation Eq. (\ref{D}) appears as the extremum of the action given by Eq. (\ref{Se0}). It is worth mentioning, that we consider the system of only one emergent Weyl spinor. In general case several emergent Weyl spinors may appear in the low energy approximation to the microscopic high energy theory. In this case the correlation between these emergent spinors may be present. As a result both the gauge field and the vierbein become matrices in flavor space. Besides, as it was mentioned above, there may appear the composite fields, that interact with the emergent Weyl spinors via the fermion number breaking terms. Such terms may be responsible for the formation of the Majorana masses of the right - handed neutrino necessary for the type I seesaw mechanism.

The considered pattern points out, that Weyl equation for the quarks and leptons of the Standard Model of particle physics may be the emergent phenomenon, which appears at low energies in the microscopic system described purely in terms of real numbers. This microscopic system contains multi - component Majorana fermions. The imaginary unit, which enters the equations for the Weyl spinors, appears effectively on the way from the microscopic coordinates $\bf Q$ to the observed coordinates $\bf X$ through momentum space that parametrizes the branches of the spectrum of the microscopic Hamiltonian.
The observed coordinates are those, in which the effective action for the emergent Weyl spinors does not contain the fermion number breaking terms (when the interaction between the Weyl spinors is neglected). The interaction between the Weyl spinors   may, in principle, contain the terms that break the fermion number.

Since Weyl fermions are "primary" objects in the Standard Model, in the effective theory  the other objects -- bosons and fermions -- are the composite objects made of Weyl fermions. That is why the wave functions of all matter fields will be also expressed via the effective complex numbers.
Thus, it is possible that the complex numbers appear together with Weyl fermions, gauge fields and gravity in the vicinity of the crossing point in the spectrum of the original microscopic system. Existence of the crossing point (Weyl point) may also solve the hierarchy problem in the Standard Model:  masses of quarks and leptons are much smaller than the characteristic Planck energy scale simply because they emerge as gapless chiral Weyl fermions.

\section*{Acknowledgements}
The work of M.A.Z. is  supported by the Natural Sciences and Engineering
Research Council of
Canada and grant RFBR 14-02-01261. GEV thanks Yu.G. Makhlin for fruitful discussions and
acknowledges a financial support of the Academy of Finland and its COE
program.



\begin{thebibliography}{99}

\bibitem{HasanKane2010}
Hasan, M.Z. and Kane, C.L.,
Topological Insulators, {\it Rev. Mod. Phys.} \textbf{82}, 3045 (2010).

\bibitem{QiZhang2010}
Qi, X.-L. and Zhang, S.-C.,
Topological insulators and superconductors,
{\it Rev. Mod. Phys.}  \textbf{83}, 1057--1110 (2011).

\bibitem{Froggatt1991}
C.D. Froggatt   and  H.B. Nielsen,
{\it Origin of Symmetry},
World Scientific, Singapore, 1991.

\bibitem{Volovik2003}
Volovik, G.E.,
{\it The Universe in a Helium Droplet},
Clarendon Press,  Oxford (2003).

\bibitem{Horava2005}
Ho\v{r}ava, P.
Stability of Fermi surfaces and $K$-theory,
{\it Phys. Rev. Lett.} \textbf{95}, 016405 (2005).

\bibitem{Adler2004}
S. L. Adler,
{\it Quantum Theory as an Emergent Phenemon: The Statistical Mechanics of Matrix Models
as the Precursor of Quantum Field Theory},
Cambridge University Press, Cambridge (2004).

\bibitem{Yang}
C.N. Yang, "Thematic Melodies of Twentieth Century
Theoretical Physics : Quantization, Symmetry and Phase Factor",
in: International Conference on
Theoretical Physics, TH-2002, Paris, July 22-27, 2002,
D. Iagolnitzer, V. Rivasseau and J. Zinn-Justin eds., Birkh\"auser Verlag,
Basel-Boston-Berlin (2004), Ann. Henri Poincare {\bf 4}, Suppl. 2,   S9--S14
(2003).


 \bibitem{Abrikosov1971}
A.A. Abrikosov    and S.D. Beneslavskii,
Possible existence of substances intermediate between metals and dielectrics,
Sov. Phys. JETP {\bf 32}, 699 (1971).

 \bibitem{Abrikosov1998}
A.A. Abrikosov,
 Quantum magnetoresistance,
 Phys. Rev. {\bf B 58}, 2788 (1998).



\bibitem{Burkov2011}
A.A. Burkov and L. Balents,
Weyl semimetal in a topological insulator multilayer,
Phys. Rev. Lett. {\bf 107}, 127205 (2011);
A.A. Burkov, M.D. Hook, L. Balents,
Topological nodal semimetals,
Phys. Rev. B {\bf 84}, 235126 (2011).

\bibitem{XiangangWan2011}
Xiangang Wan, A.M. Turner,  A. Vishwanath  and S.Y. Savrasov,
 Topological semimetal and Fermi-arc surface states in the electronic
structure of pyrochlore iridates,
Phys. Rev. B {\bf 83}, 205101 (2011).

\bibitem{Weylsemimetal}
S. Borisenko, Q. Gibson, D. Evtushinsky, V. Zabolotnyy, B. Buechner, R.J. Cava,
Experimental Realization of a Three-Dimensional Dirac Semimetal,
 arXiv:1309.7978 [cond-mat.mes-hall]

\bibitem{Semenoff:1984dq}
  G.~W.~Semenoff,
  Condensed Matter Simulation Of A Three-dimensional Anomaly,
{ Phys.\ Rev.\ Lett.}\  {\bf 53}, 2449 (1984).


\bibitem{VolkovPankratov1985}
B.A. Volkov and O.A. Pankratov,
Two-dimensional massless electrons in an inverted contact,
JETP Lett. {\bf 42}, 178--181 (1985).


\bibitem{HasanKane2010}
M.Z. Hasan and C.L. Kane,
Topological Insulators,
Rev. Mod. Phys. {\bf 82}, 3045--3067 (2010).

\bibitem{Xiao-LiangQi2011}
Xiao-Liang Qi and Shou-Cheng Zhang,
Topological insulators and superconductors,
Rev. Mod. Phys. {\bf 83}, 1057--1110 (2011).

\bibitem{SalomaaVolovik1988}
 M.M. Salomaa and  G.E. Volovik,
 Cosmiclike domain walls in superfluid $^3$He-B: Instantons and diabolical points in (${\bf k}$,${\bf r}$) space, Phys. Rev.  {\bf B~37}, 9298--9311 (1988).


\bibitem{Volovik2009}
 G.E. Volovik,
Fermion zero modes at the boundary of superfluid $^3$He-B,
 Pis'ma ZhETF {\bf 90}, 440--442 (2009); JETP Lett. {\bf 90}, 398--401 (2009);
arXiv:0907.5389.


 \bibitem{NeumannWigner}
J. von Neumann und E. Wigner,
 \"Uber merkw\"urdige diskrete Eigenwerte, Z. Phys. {\bf 30}, 465--467
(1929);
J. von Neumann  und E.P. Wigner,
\"Uber das Verhalten von Eigenwerten bei adiabatischen Prozessen,
 Phys. Zeit. {\bf 30}, 467--470 (1929).

\bibitem{Novikov1981}
S.P. Novikov,
 Magnetic Bloch functions and vector bundles. Typical dispersion laws and their quantum numbers,
 Sov. Math., Dokl. {\bf 23}, 298--303 (1981).


\bibitem{Avron1983}
J.E. Avron, R. Seiler and B. Simon,
Homotopy and quantization in condensed matter physics,
Phys. Rev. Lett. {\bf 51}, 51--53 (1983).

\bibitem{Volovik1987}
G.E. Volovik,
Zeros in the fermion spectrum  in superfluid systems as diabolical points,
Pis'ma ZhETF { \bf 46}, 81--84 (1987); JETP Lett. {\bf 46}, 98--102 (1987).

\bibitem{Volovik1986A}
G.E. Volovik,
Analog of gravity in superfluid  $^3$He-A,
JETP Lett. {\bf 44}, 498--501 (1986).

\bibitem{teleparallel}
K.Hayashi, T.Shirafuji, Phys. Rev. D19, 3524 (1979)
Yu. N. Obukhov, J. G. Pereira, Phys.Rev. D67 (2003) 044016
V. C. de Andrade, J. G. Pereira, Phys.Rev. D56 (1997) 4689-4695


\bibitem{VK2010} M. A. H. Vozmediano, M. I. Katsnelson, and F. Guinea,
Phys. Rep. 496 109(2010).

\bibitem{Manes2013}
Juan L. Manes, Fernando de Juan, Mauricio Sturla, Mar�a A. H. Vozmediano,
 Generalized effective hamiltonian for graphene under non-uniform strain,
Phys. Rev. B {\bf 88}, 155405 (2013); arXiv:1308.1595.

\bibitem{VZ2014_NPB}
 G.E.Volovik, M.~A.~Zubkov,
 { ``Emergent Weyl spinors in multi-fermion systems,''}
   Nuclear Physics B
Volume 881, April 2014, Pages 514 $-–$ 538, DOI:	10.1016/j.nuclphysb.2014.02.018,
  arXiv:1402.5700.


\bibitem{ZV2013}
G.~E.~Volovik and M.~A.~Zubkov,
  ``Emergent Horava gravity in graphene,''
  arXiv:1305.4665 [cond-mat.mes-hall], Annals of Physics 340/1 (2014), pp. 352-368, doi:10.1016/j.aop.2013.11.003


 \bibitem{Volovik2011}
 G.E. Volovik,
The topology of quantum vacuum,
in:
Analogue Gravity Phenomenology,
Analogue Spacetimes and Horizons, from Theory to Experiment,
 Lecture Notes in Physics,  {\bf 870}, 343--383 (2013),
Faccio, D.; Belgiorno, F.; Cacciatori, S.; Gorini, V.; Liberati, S.; Moschella, U. (Eds.);
 arXiv:1111.4627.

\end{thebibliography}
\end{document}